\documentclass[superscriptaddress,pre,twocolumn]{revtex4}
\usepackage{epsfig,amsmath,amssymb,graphics,color}
\newcommand{\be}{\begin{equation}}
\newcommand{\ee}{\end{equation}}

\newcommand{\ba}{\begin{eqnarray}}
\newcommand{\ea}{\end{eqnarray}}

\begin{document}

\title{
Direct experimental evidence of a growing
length scale\\ accompanying the glass transition
}

\author{L.~Berthier}
\affiliation{Laboratoire des Collo{\"\i}des, Verres
 et Nanomat{\'e}riaux UMR 5587, Universit{\'e} Montpellier II and CNRS,
34095 Montpellier, France}

\author{G.~Biroli}
\affiliation{Service de Physique Th{{\'e}o}rique
Orme des Merisiers -- CEA Saclay, 91191 Gif sur Yvette Cedex, France}

\author{J.-P.~Bouchaud}
\affiliation{Service de Physique de l'{\'E}tat Condens{\'e}
Orme des Merisiers -- CEA Saclay, 91191 Gif sur Yvette Cedex, France}
\affiliation{Science \& Finance, Capital Fund Management
6-8 Bd Haussmann, 75009 Paris, France}

\author{L.~Cipelletti}
\affiliation{Laboratoire des Collo{\"\i}des, Verres
 et Nanomat{\'e}riaux UMR 5587, Universit{\'e} Montpellier II and CNRS,
34095 Montpellier, France}

\author{D.~El Masri}
\affiliation{Laboratoire des Collo{\"\i}des, Verres
 et Nanomat{\'e}riaux UMR 5587, Universit{\'e} Montpellier II and CNRS,
34095 Montpellier, France}

\author{D.~L'H{\^o}te}
\affiliation{Service de Physique de l'{\'E}tat Condens{\'e}
Orme des Merisiers -- CEA Saclay, 91191 Gif sur Yvette Cedex, France}

\author{F.~Ladieu}
\affiliation{Service de Physique de l'{\'E}tat Condens{\'e}
Orme des Merisiers -- CEA Saclay, 91191 Gif sur Yvette Cedex, France}

\author{M.~Pierno}
\affiliation{Laboratoire des Collo{\"\i}des, Verres
 et Nanomat{\'e}riaux UMR 5587, Universit{\'e} Montpellier II and CNRS,
34095 Montpellier, France}

\date{\today}

\begin{abstract}
{\bf Understanding glass formation is a challenge because the existence of
a true glass state, distinct from liquid and solid, remains elusive: Glasses
are liquids that have become too viscous to flow. An old idea, as yet
unproven experimentally, is that the dynamics becomes sluggish
as the glass transition approaches
because increasingly larger
regions of the material have to move simultaneously to allow flow.
We introduce new multipoint dynamical susceptibilities to
estimate quantitatively the size of these regions and provide
direct experimental evidence that the glass formation of
molecular liquids and colloidal suspensions is accompanied by growing
dynamic correlation length scales.
}
\end{abstract}

\maketitle

Why does the viscosity of glass-forming liquids increase so
dramatically when approaching the glass transition? Despite
decades of research a clear explanation of this phenomenon, common
to materials as diverse as molecular glasses, polymers, and
colloids is still lacking~\cite{Donth,DS}. The conundrum is that
the static structure of a glass is indistinguishable from that of
the corresponding liquid, with no sign of increasing static
correlation length scales accompanying the glass transition. Numerical
simulations performed well above the glass temperature, $T_g$,
reveal instead the existence of a growing dynamic
length scale~\cite{harrowell,glotzer,glotzer2,steve,berthier}
associated to dynamic heterogeneities~\cite{ediger}.
Experiments~\cite{ediger,exp0,exp1,exp2,exp3} have indirectly
suggested a characteristic length scale of about 5 to 20 molecular diameters at
$T_g$, but its time and temperature dependencies, which are crucial 
for relating this finding to the glass transition, were not
determined.

We present quantitative
experimental evidence that glass formation in molecular liquids
and colloids is accompanied by at least one growing dynamic
length scale. We introduce experimentally accessible
multipoint dynamic susceptibilities that quantify the correlated
nature of the dynamics in glass formers. Because 
these measurements can be made using a wide
variety of techniques in vastly different materials,
a detailed characterization of the microscopic mechanisms
governing the formation of amorphous glassy states becomes possible.

Supercooled liquids are believed to exhibit spatially
heterogeneous dynamics over
length scales that grow when approaching the glass
state~\cite{Donth,Wolynes,Gilles,gc}. This
heterogeneity implies the existence of significant fluctuations
of the dynamics because the  number of independently
relaxing regions is reduced.
Numerical simulations have focused on
a ``four-point'' dynamic susceptibility $\chi_{4}(t)$,
which quantifies the amplitude of spontaneous fluctuations around the
average dynamics~\cite{harrowell,glotzer,glotzer2,steve,berthier}.
The latter is usually measured through ensemble-averaged
correlators, $F(t) = \langle \delta A (t) \delta A(0) \rangle
= \langle C(t) \rangle$, where $\delta
A(t) = A(t) - \langle A \rangle$ represents the
spontaneous fluctuation of an observable $A(t)$, such as the density.
Dynamic correlation leads to large fluctuations of $C(t)$,
measured by $\chi_4(t) = N
\langle \delta C^2(t) \rangle$, where $N$ is the
number of particles in the system.
The susceptibility $\chi_4(t)$ typically presents a nonmonotonic
time dependence with a peak centered at the liquid's relaxation
time~\cite{TWBBB}.
The height of this peak is
proportional to the volume within which
correlated motion takes place~\cite{glotzer,glotzer2,gc,TWBBB}.
Unfortunately, numerical
findings are limited to short timescales ($\sim 10^{-7}$ s)
and temperatures far above $T_g$.
Experimentally, detecting spontaneous fluctuations of dynamic
correlators remains an open challenge, because dynamic
measurements have to be resolved in both  space and time~\cite{mayer}.

Induced fluctuations are
more easily accessible experimentally than spontaneous ones and
can be related to one another by fluctuation-dissipation theorems.
We introduce a dynamic susceptibility defined as the response
of the correlator $F(t)$ to a perturbing field $x$: 
\begin{equation}
\chi_x (t) = \frac{\partial F(t)}{\partial x}
\label{un}
\end{equation} 
The relaxation time of supercooled liquids increases
abruptly upon cooling, so a relevant perturbing field is temperature,
in which case
Eq.~\ref{un} becomes $\chi_T (t) = \partial F(t) /
\partial T$. Density also plays a role in supercooled liquids,
although a less crucial one~\cite{alba}. Hence, another
interesting susceptibility is $\chi_P(t)=\partial F(t) /
\partial P$, where $P$ is the pressure.
Colloidal hard spheres undergo a glass transition~\cite{pusey} at
high particle volume fraction $\varphi$. Thus, the appropriate
susceptibility for colloids is $\chi_\varphi (t) = \partial F(t)
/\partial \varphi$. Equation~\ref{un} also applies in the
frequency domain, $\chi_x(\omega) =
\partial {\tilde F}(\omega)/
\partial x$, where ${\tilde F}(\omega)$ can be the
dielectric susceptibility. We will show below that linear response
formalism and fluctuation theory can be used to relate
$\chi_x(t)$ to the spontaneous fluctuations of $C(t)$, and thus
to $\chi_4(t)$. Thus, $\chi_x(t)$ is an experimentally
accessible multi-point dynamic susceptibility that directly
quantifies dynamic heterogeneity in glass formers.

For molecular liquids, the dynamics conserves energy, volume and
number $N$ of particles, and one can establish, in the $NPT$ ensemble
relevant for experiments,
the following fluctuation-dissipation theorem, 
\begin{equation} k_{B} T^2
\chi_T(t) = N \langle \delta C (t) \delta H (0) \rangle
\label{fdt} 
\end{equation} where $k_{B}$ is the Boltzmann constant, $H(t)$
the fluctuating enthalpy per particle, and $C(t)$  the
instantaneous value of a generic dynamic correlator $F(t)$. Both
$C(t)$ and $H(t)$ are sums over local contributions~\cite{hansen},
$NC(t)=\rho\int d^3 \vec r c(\vec r,t)$ and $NH(t) = \rho
\sqrt{k_B c_P} T \int d^3 \vec r {\hat h}(\vec r,t)$. Here, $\rho$
is the average number density, $c_P$ the constant pressure specific heat
that sets the scale of the enthalpy fluctuations, $\langle \delta
H^2 \rangle= k_B c_P T^2$, so that the field ${\hat h}(\vec r,t)$
has unit variance. Using translational invariance, Eq.~\ref{fdt}
can be rewritten as: 
\begin{equation} 
\sqrt{\frac{k_B}{c_P}} T \chi_T(t)= \rho
\int d^3 \vec r \left\langle \delta c(\vec r,t) \delta {\hat
h}(\vec 0,0) \right\rangle 
\label{length} 
\end{equation} 
This expression shows that $\chi_T(t)$ directly probes the range of
spatial correlations between local fluctuations of the dynamics
and that of the enthalpy. In the case of colloids, the
dynamics only conserves density and a similar expression can be
obtained 
\begin{equation} 
\sqrt{\rho k_B T \kappa_T} \, \varphi \chi_\varphi(t)
= \rho \int d^3 \vec r \left\langle \delta c(\vec r,t) \delta \hat
\rho(\vec 0,0) \right\rangle 
\label{lengthrho} 
\end{equation} 
where
$\kappa_T$ is the isothermal compressibility and $\delta \hat
\rho$ denotes density fluctuations rescaled by their root mean
square.

Equations \ref{length} and \ref{lengthrho} show that
$\chi_x(t)$ probes the extent of spatial dynamic correlations that
differ from the ones studied in earlier theoretical and numerical
works, which focused instead on $\chi_4(t)= \rho \int d^3 \vec r
\langle \delta c(\vec r,t) \delta c(\vec 0,t) \rangle$. We have,
however, established a direct relation between $\chi_x(t)$ and
$\chi_4(t)$ by using the thermodynamic formalism developed in
\cite{lebo}, which is generically applicable to bulk
glass formers above the glass transition. For dynamics conserving
energy and volume, we relate the fluctuations of $C(t)$, and
therefore $\chi_4(t)$ measured in the $NPT$ ensemble, to its
isobaric-isoenthalpic counterpart, $\chi_4^{NPH}(t)$, which
quantifies the amplitude of the fluctuations of $C(t)$ in the
$NPH$ ensemble in which all configurations have exactly the same
enthalpy: $\chi_4(t) = \chi_4^{NPH}(t) + k_B T^2
\chi_T^2(t) / c_P$. Because $\chi_4^{NPH}(t) > 0$, one derives
an experimentally measurable rigorous lower bound~\cite{toto}
 for $\chi_4(t)$:
\begin{equation}
\label{ineqT}
\chi_4(t) \geq \frac{k_B}{c_P} T^2 \chi_T^2(t)
\end{equation}
A similar inequality holds between $ \chi_4(\omega)$ and
$\chi_T(\omega)$, where $\chi_4(\omega)$ denotes the amplitude of
spontaneous fluctuations around ${\tilde F}(\omega)$. Similar
arguments also apply to $\chi_4(t)$ computed in the $NVT$ ensemble
preferred in numerical simulations. In that case, energy
replaces enthalpy in Eqs.~\ref{fdt} and \ref{length}, and the
specific heat at constant volume, $c_V$, replaces $c_P$ in
Eq.~\ref{length} and relation \ref{ineqT}. Finally, we find that an
inequality similar to relation \ref{ineqT} holds for colloidal systems,
for which the volume and the number of particles are conserved
quantities: 
\begin{equation} 
\label{ineqrho} 
\chi_4 (t) \geq \rho k_{B}T
\kappa_T \, \varphi^2 \chi_\varphi^2(t) 
\end{equation}

We have determined $\chi_x(t)$ experimentally and numerically in
three representative glass formers. For supercooled glycerol near
$T_g \approx 185$~K, the real part of the dielectric susceptibility,
$\epsilon'(\omega)$, was measured every 1 K in the 
temperature range from 192~K to 232~K. 
After fitting to a Havriliak-Negami form~\cite{Donth}, we use
smoothed finite differences to evaluate $\chi_T(\omega) =
\partial [\epsilon'(\omega)/\epsilon'(0)]/\partial T$ and
show in Fig.~1A the right hand side
of relation \ref{ineqT} as a function of inverse frequency. We plot in
Fig.~1B the right side of relation \ref{ineqrho} for hard
sphere colloids where $\chi_\varphi (t) = \partial f(q,t) /
\partial \varphi$. The normalized intermediate scattering
function~\cite{hansen} $f(q,t)$ is measured by dynamic light
scattering~\cite{djamel} for a wavevector $q$ close to the first
peak of the static structure factor. Several packing fractions are
studied, from diluted samples where $f(q,t)$ decays exponentially
in $\sim 1$ ms to concentrated suspensions with a two-step decay
and a final relaxation time of $\sim 10$ s. Finite differences of
data sets obtained for nearby $\varphi$ are used to deduce
$\chi_{\varphi}(t)$. Finally, we show in Fig.~1C
numerical data obtained by standard molecular dynamics simulations
of a binary Lennard-Jones mixture, a well-studied model for
fragile supercooled liquids~\cite{KA,KAbis}. The dynamics is recorded at
nearby temperatures through the self part of the intermediate
scattering function, whose characteristic decay time spans a
range from 1 ps to 100 ns (using Argon units~\cite{KA,KAbis}).

\begin{figure}
\psfig{file=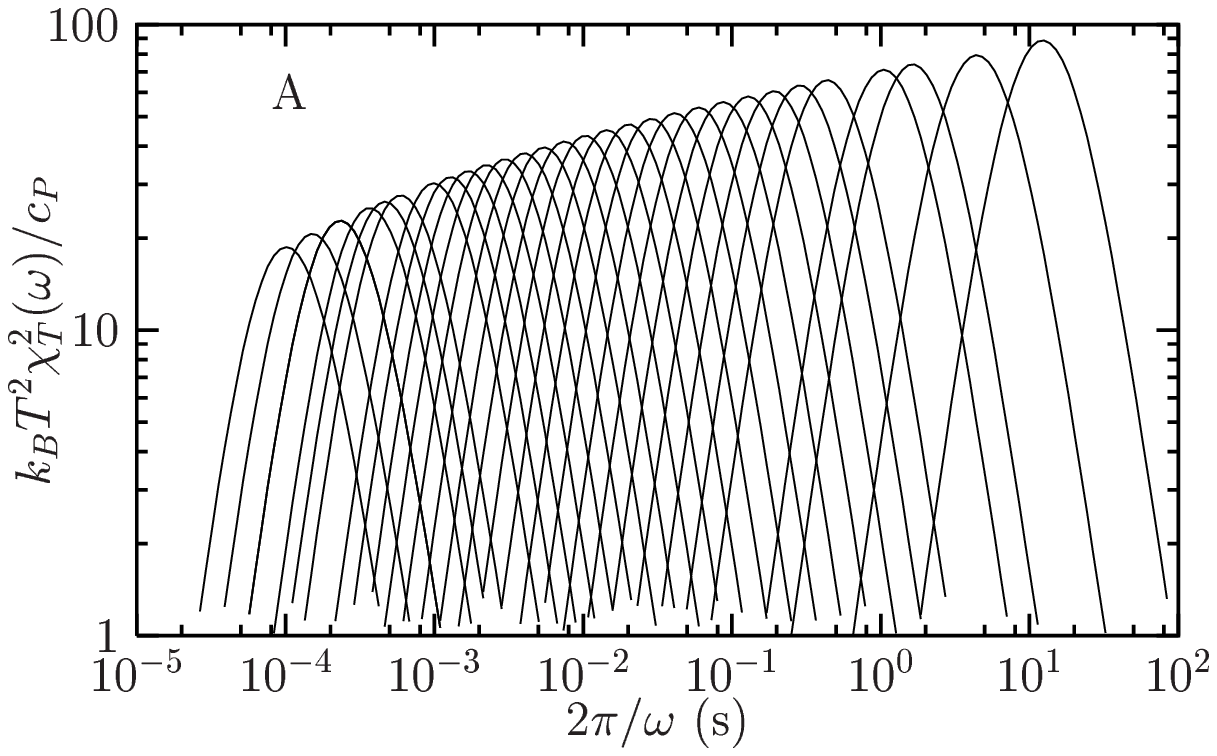,width=8.5cm}
\psfig{file=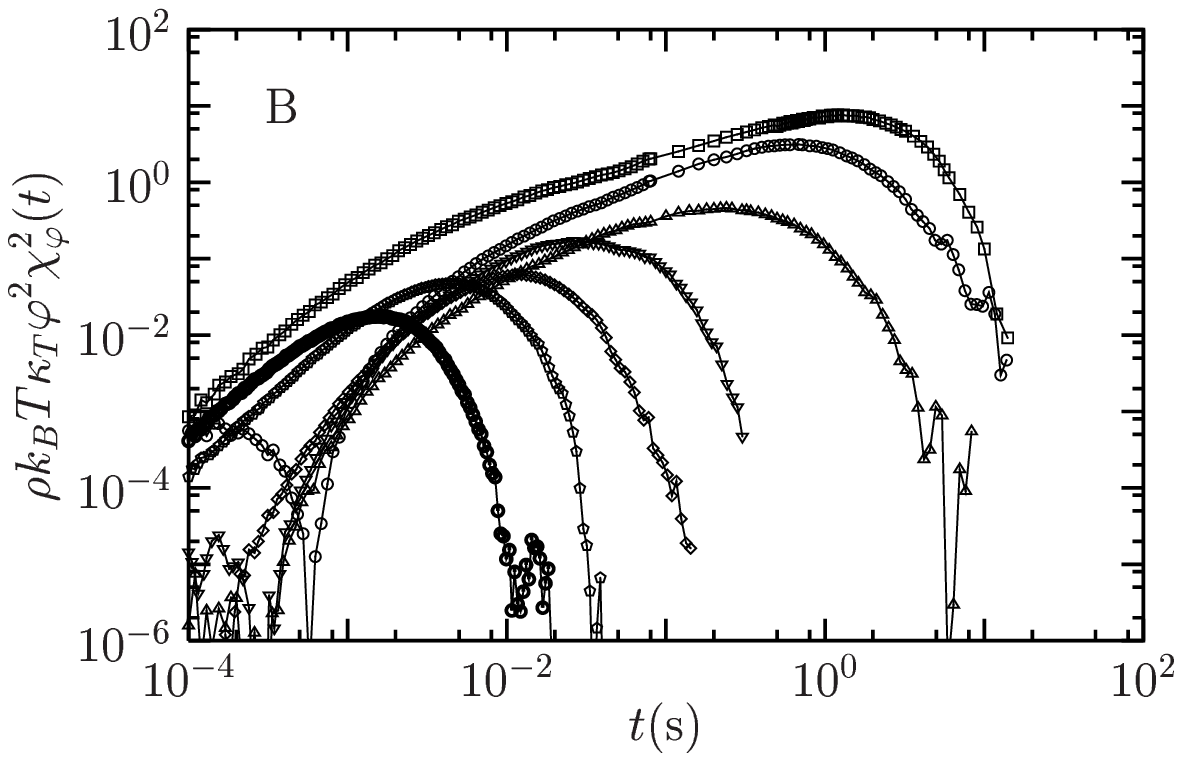,width=8.5cm}
\psfig{file=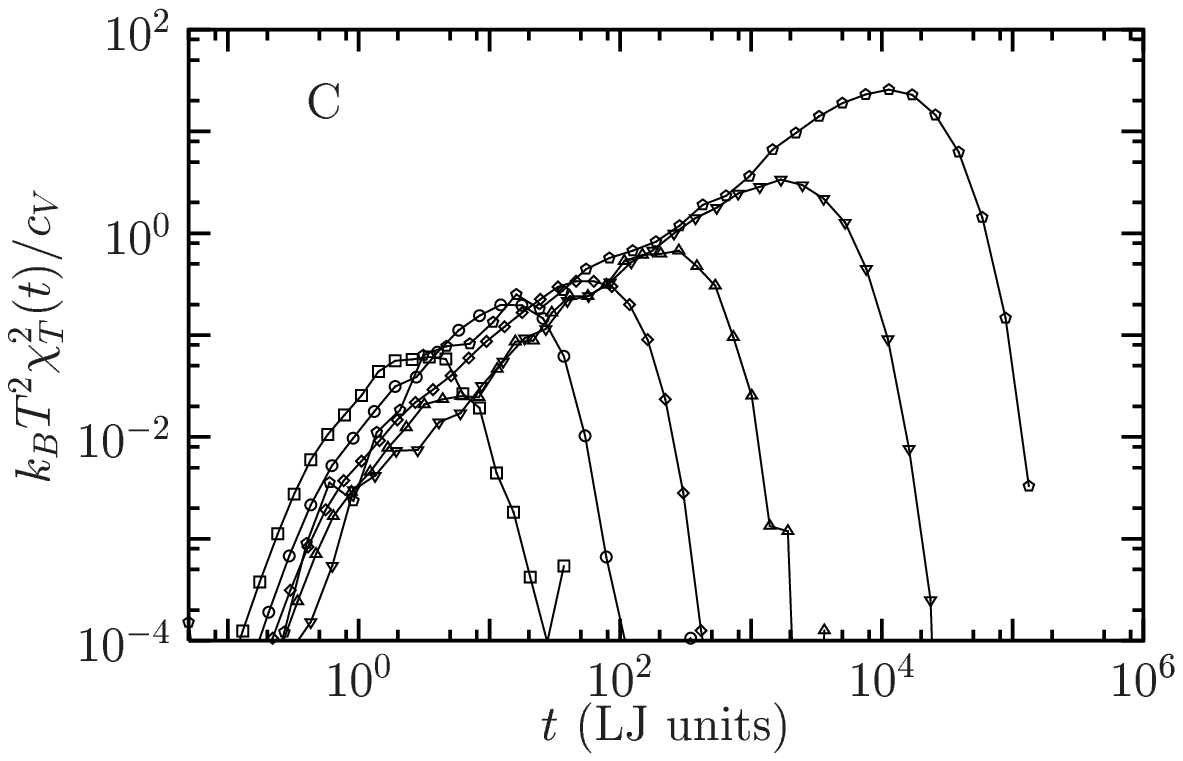,width=8.5cm}
\caption{Dynamic
susceptibilities in ``$\chi_4$ units'', right side of
relations \ref{ineqT} and \ref{ineqrho} for three glass-formers.
(A)
$\chi_T(\omega)$ was obtained for 99.6\% pure supercooled glycerol
(in a desiccated Argon environment to prevent water absorption) by
using standard capacitive dielectric measurements for $192
~\textrm{K} \leq T \leq 232 ~\textrm{K}$ ($T_g \approx 185$ K). 
(B) $\chi_\varphi(t)$ was obtained in colloidal hard spheres via
dynamic light scattering. The static prefactor, $\rho k_B T
\kappa_T$, was evaluated from the Carnahan-Starling equation of
state~\cite{hansen}. From left to right, $\varphi=0.18$, 0.34,
0.42, 0.46, 0.49, and 0.50. 
(C) $\chi_T(t)$ was obtained in a
binary Lennard-Jones (LJ) mixture via numerical simulations. From
left to right, $T=2.0$, 1.0, 0.74, 0.6, 0.5 and 0.465 (in reduced
LJ units~\cite{KA,KAbis}). Relative errors at the peak
are at most about 10\% for (A) and (C), and 30\% for (B).
For all of the systems, dynamic susceptibilities
display a peak at the average relaxation time whose height
increases when the dynamics slows down, which is direct evidence of
enhanced dynamic fluctuations and a growing dynamic length scale.}
\end{figure}

Dynamical susceptibilities behave similarly in all three cases.
All display a peak for $t \approx \tau_\alpha$, the average
relaxation time. The peak height increases when the glass
transition is approached. This behavior represents the central
result of our work. Together with Eqs.~\ref{length} to
\ref{ineqrho}, it provides direct evidence of enhanced dynamic
fluctuations and a growing dynamic lengthscale associated with the
glass transition.

How tight the bounds of relations \ref{ineqT} and \ref{ineqrho} 
are depends upon
the specific material and range of parameters studied. A
quantitative answer is given by simulations where the 
microcanonical quantity
$\chi_4^{\rm micro}(t)$,  i.e. the difference
between $\chi_4(t)$ in the $NVT$ ensemble and $k_B T^2\chi_T(t)/c_V$,
can be easily measured.
For the Lennard-Jones mixture, we find that the right
side of relation \ref{ineqT} is much smaller
than $\chi_4(t)$ at high $T$, but the difference 
rapidly diminishes when $T$
decreases. Both sides of relation \ref{ineqT} become comparable for the
lowest temperature shown in Fig.~1C, which is still well
above $T_g$. Following Ref.~\cite{BB}, we also find that
mode-coupling theory predicts $\chi_4(t) \sim \chi_T^2(t) \sim
(T/T_c-1)^{-2}$ near the mode-coupling singularity $T_c >
T_g$, provided that conserved variables are properly taken into
account. 

These results support the idea that relation \ref{ineqT} can
be used as an equality to quantitatively estimate $\chi_4(t)$ at low
temperature, at least for fragile systems. This use of 
relation \ref{ineqT} is equivalent to
assuming that dynamic heterogeneity in molecular liquids is
strongly correlated with enthalpy fluctuations, and through
a similar argument, with density fluctuations in colloids.
In fact, supposing that
enthalpy is the only source of fluctuations,
$\delta C \approx (\partial C/ \partial T)_P \delta H/c_P$, and
if we use the definition of $c_P$, we obtain directly 
that $\chi_4(t) \approx
k_B T^2 \chi_T^2(t) / c_P$. A more
general result can be obtained by taking into account that energy
and density are both fluctuating quantities, in which case
$\chi_4(t)$ is the sum of two contributions: $\chi_4(t) \approx k_B
T^2 \chi_T^2(t)/c_V + \rho k_B T \kappa_T \rho^2 \chi_\rho^2(t)$.
The second term is negligible in most fragile liquids
\cite{alba}, but dominates in colloidal systems. The presence of
additional sources of fluctuations justifies that the rigorously
derived inequality \ref{ineqT} does not hold as an equality.

Our results for dynamic fluctuations provide an estimate of the
size $\xi$ of dynamic heterogeneity in liquids near $T_g$.
Because $\chi_4 (t) = \rho \int d^3\vec r \langle \delta c(\vec r,t)
\delta c(\vec 0,t) \rangle$, this quantity, once divided 
by the amplitude of the fluctuations at zero distance,
$\langle \delta c^2({\vec 0},t) \rangle$, 
defines a correlation volume.
The correlation functions are normalized to unity at
$t=0$, so $\langle \delta c^2({\vec 0},t) \rangle$ is of order one or
smaller. Our simulations indeed show that in the temperature
regime where the dynamics slows down and on timescales not much
longer than the system relaxation time, this average is of the
order of one and displays extremely weak temperature dependence, as expected
for a local quantity in glass formers. Thus, the height of the
peak in the dynamic susceptibility, $\chi_4^\star\approx k_B T^2
(\chi_T^\star)^2/c_P$, yields directly a correlation volume
expressed in molecular units, $\chi_4^\star \simeq
(\xi/a)^{\zeta}$, where $a$ is the molecular size and $c_P$ is
expressed in units of $k_B$. Numerical~\cite{glotzer,steve} and
theoretical~\cite{steve,gc,TWBBB,BB} works suggest that $\zeta
\approx 2$ to 4. 

A direct comparison between our data and existing
measurements can be performed for glycerol, where multidimensional 
nuclear magnetic resonance (NMR)
experiments show that $\xi = 1.3 \pm 0.5$ nm for 
$T=199$~K \cite{mark,encoremark}. Assuming a simple compact geometry for
heterogeneities, $\zeta = 3$, we estimate that $\xi$ increases
from $0.9$ nm at $T=232$ K to $1.5$ nm at $192$ K. Given the
assumptions involved in both approaches, and the uncertainty about
numerical prefactors of order unity, the agreement is remarkable.
An important physical conclusion of
our work is that dynamic heterogeneity is strongly correlated to enthalpy
fluctuations in fragile liquids,  although  there is no
signature of any static large-scale correlations~\cite{harrowell,steve,gc}.

For other glass-forming liquids, we obtain an
estimate for $\xi$ at $T_g$ by assuming for
simplicity that correlators obey time-temperature superposition,
$F(t) = {\cal F}(t/\tau_\alpha)$, and using relation 
\ref{ineqT} as an
equality. One gets 
\begin{equation} 
\label{peak} \chi_4^\star(T_g) \approx
[{\cal F}'(1)]^2 \frac{k_B}{c_P} \left(\frac{\partial \ln
\tau_\alpha}{\partial \ln T} {\Big \vert}_{T_g} \right)^2
\end{equation} 
The logarithmic derivative is proportional to the 
well-known ``steepness index'' $m$,
introduced in the glass literature to characterize the
fragility of glass-forming liquids~\cite{WangAngell}. From 
reported values~\cite{Donth,WangAngell} of the quantities
appearing in Eq.~\ref{peak} 
and assuming a stretched exponential form 
for ${\cal F}(x)= \exp(-x^\beta)$,  
we estimate $\chi_4^\star(T_g)$ for
different glass-forming liquids in Fig.~2A. 

\begin{figure}
\psfig{file=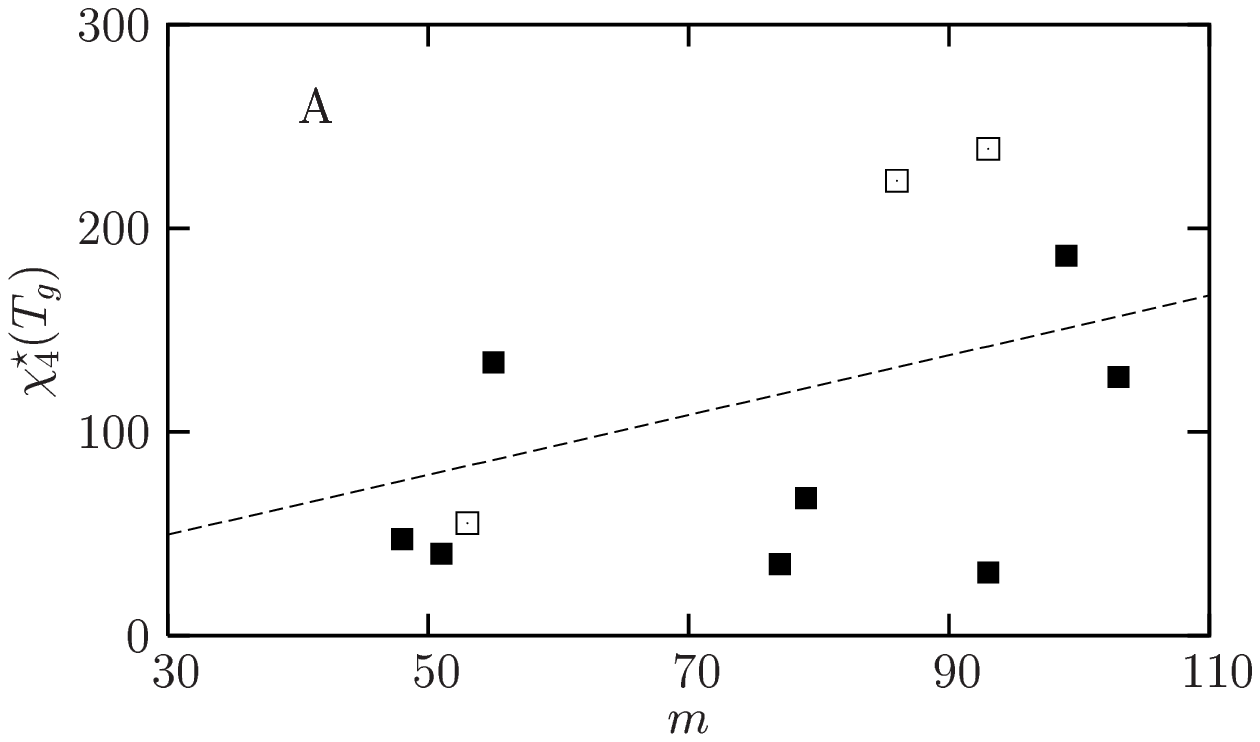,width=8.5cm}
\psfig{file=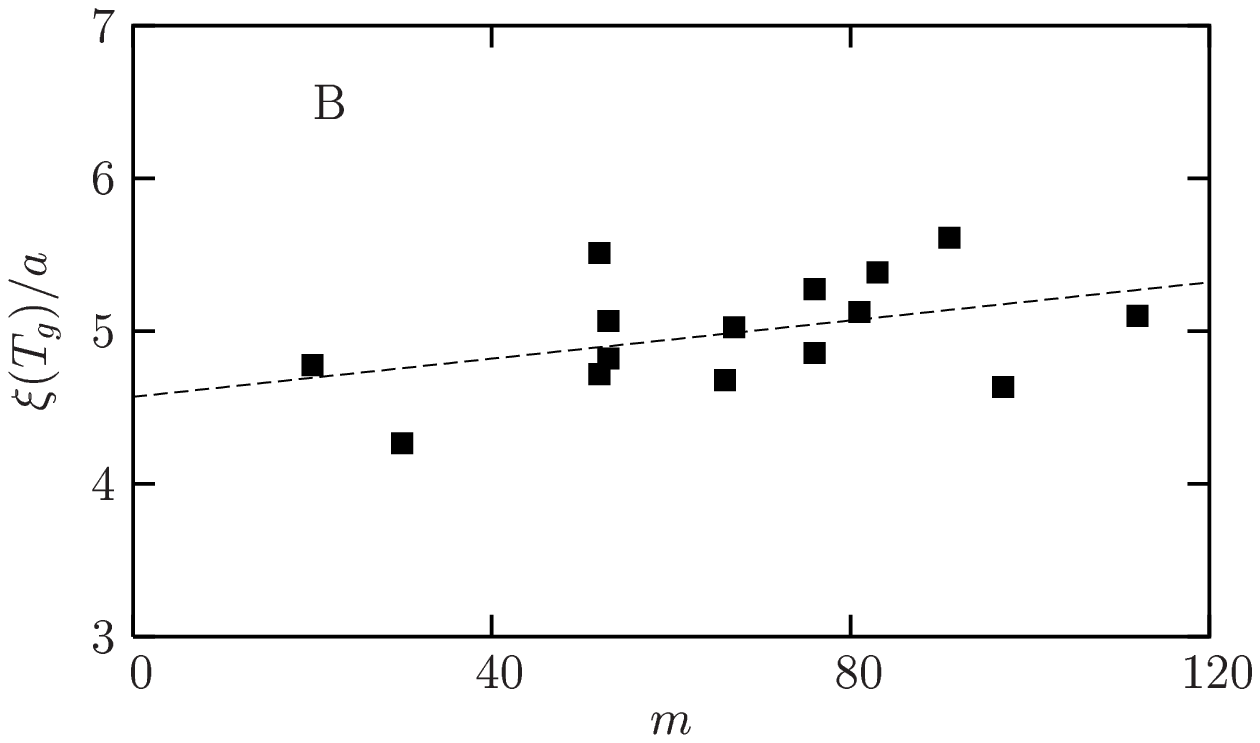,width=8.5cm}
\caption{(A) Correlation volume $\chi_4^\star(T_g)$
in supercooled liquids at the glass transition. 
Filled squares represent a lower bound to $\chi_4^\star(T_g)$ 
in molecular units estimated
through Eq.~\ref{peak}. Different points represent different
materials, which are ranked by their fragility $m$.
Open squares represent the same quantity evaluated 
from available multidimensional NMR data,
using~\cite{encoremark} $\chi_4^\star = \int d^3 r \exp(-2r/\xi)$,  
for  glycerol ($T_g$+ 10 K)~\cite{mark}, 
orthoterphenyl ($T_g$ + 9 K)~\cite{encoremark,otp}
and d-sorbitol ($T_g$+ 7 K)~\cite{encoremark}.
A linear fit to the weak increase of $\chi_4^\star$ with fragility is 
shown as a dashed line.
(B) Correlation length $\xi(T_g)$
in supercooled liquids at the glass transition expressed in 
bead units $a$.
The correlation volume is first evaluated 
using $\Delta c_P$ instead of $c_P$ in
Eq.~\ref{peak}. Following~\cite{StevensonWolynes}, $\Delta c_P$ is expressed
in $k_B$ ``per bead'' units accounting for different molecular
shapes and sizes.
Using $\zeta=3$, the result is finally 
converted into a length scale expressed in bead
units.
The known empirical correlations~\cite{WangAngell} between $m$, $\beta$ 
and $\Delta c_P$ translate into a weak 
increase of $\xi(T_g)$ with fragility, which we fit with a linear 
relation shown as a dashed line.}
\end{figure}

For complex molecules, fluctuations
that are unrelated to the glassy dynamics might contribute to the
specific heat. These effects may be taken into account by replacing $c_P$
in Eq.~\ref{peak} by $\Delta c_P$, the jump in specific heat at
$T_g$, which is sensitive only to the glassy degrees of freedom.
Furthermore, for large molecules, the molecular size is probably
not the relevant microscopic lengthscale, and it is sensible to
express the specific heat in units of $k_B$ per ``independent
bead'' instead of molecular units~\cite{StevensonWolynes}.
These physical assumptions are used in Fig.~2B, 
where we have converted our results into lengthscales
expressed in bead units, and they 
lead to a trend similar to that of the main plot but with less scatter:
Dynamic correlations revealed by $\chi_T$ increase weakly 
with fragility~\cite{encoredonth}. 
This result is compatible with some theoretical 
approach~\cite{rfot}, but 
contrasts with others that predict an opposite trend~\cite{gc2}.
This discrepancy might arise from the existence of at least two
physically distinct dynamic lengthscales, one revealed by
$\chi_T$, and a second associated to $\chi_4$. Although we found that
both quantities are comparable for fragile systems, the bound
in relation \ref{ineqT} may underestimate $\chi_4$ for strong materials.

To further test our length scale estimate of Eq.~\ref{peak}, we apply
the formula to a polymeric liquid poly(vinyl acetate) (PVAc)
using the monomer size for $a$~\cite{mark}. We  
find $\xi \approx 2.0$ nm at $T_{g}$, to be
compared with the value of $3.7 \pm 1$ nm obtained at $T_{g}+10K$
\cite{exp0,encoremark} [we assume $\zeta =3$, and
use the data on PVAc given in 
\cite{exp0,encoremark}]. Again, the agreement is
satisfactory.
A similar agreement is found
for orthoterphenyl and sorbitol, for which available NMR data
are reported in Fig.~2A.
Hence, we find that typical values for the
dynamic correlation length at $T_g$ obtained via Eq.~\ref{peak} are
in good agreement with previous experiments performed near
$T_g$~\cite{Donth,ediger,exp0,mark,otp}. However, our
approach has a broader scope, because it allows one to extend
experimental studies of dynamic heterogeneity to a range of
temperatures not previously accessible and to the full time
dependence of the fluctuations (Fig.~1).
Finally, we remark that even for (strong) Arrhenius molecular
liquids with activation energy $E$, 
relation \ref{ineqT} and time temperature superposition give
$\chi_4^\star(T) \ge (k_B/c_P) \times E^2/(k_B T)^2$, showing that dynamic
heterogeneity  must also exist in that case~\cite{tutu}, in agreement
with the general argument that for systems with finite range
interactions, diverging time scales must be accompanied by
diverging length scales.

Our experiments provide a quantitative demonstration that dynamic
correlations and length scales increase as the glass transition is
approached. More work is needed to characterize the time and
temperature dependencies of dynamic fluctuations over a larger
range of materials and parameters. Open issues also concern the
precise space-time geometry of dynamic heterogeneity that fixes
the value of the exponent $\zeta$ and the relation between
time scales and length scales, the connection between cooperativity and 
heterogeneity, and the extension of our results to
the nonequilibrium aging dynamics encountered in the glass phase.


\begin{thebibliography}{99}

\bibitem{Donth} E. Donth, {\it The glass transition} 
(Springer, Berlin, 2001).

\bibitem{DS} P.G. Debenedetti and F.H. Stillinger,
{\it Nature} {\bf 410}, 259 (2001).

\bibitem{harrowell} M.M. Hurley and P. Harrowell,
{\it Phys. Rev. E} {\bf 52}, 1694 (1995).

\bibitem{glotzer}
C. Bennemann, C. Donati, J. Baschnagel, and S.C. Glotzer,
{\it Nature} {\bf 399}, 246 (1999).

\bibitem{glotzer2}
C. Donati, S. Franz, S.C. Glotzer, G. Parisi,
{\it J. Non-Cryst. Solids} {\bf 307-310}, 215 (2002).

\bibitem{steve} 
S. Whitelam, L. Berthier, J.P. Garrahan,
{\it Phys. Rev. Lett.} {\bf 92}, 185705 (2004).

\bibitem{berthier} 
L. Berthier,
{\it Phys. Rev. E} {\bf  69}, 020201 (2004).

\bibitem{ediger} 
M.D. Ediger,
{\it Annu. Rev. Phys. Chem.} {\bf 51}, 99 (2000).

\bibitem{exp0} U. Tracht, {\it et al.},
{\it Phys. Rev. Lett.} {\bf 81}, 2727 (1998).

\bibitem{exp1} E. Weeks, J.C. Crocker, A.C. Levitt,
A. Schofield, D.A. Weitz,
{\it Science} {\bf 287}, 627 (2000).

\bibitem{exp2}
E. Vidal-Russell and N.E. Israeloff,
{\it Nature} {\bf  408}, 695 (2000).

\bibitem{exp3}
L.A. Deschenes and D.A. Vanden Bout,
{\it Science} {\bf 292}, 255 (2001).

\bibitem{Wolynes} T.R. Kirkpatrick, D. Thirumalai,
and P.G. Wolynes,
{\it Phys. Rev. A} {\bf 40}, 1045 (1989).

\bibitem{Gilles} P. Viot., G. Tarjus, and D. Kivelson, 
{\it J. Chem. Phys.} {\bf 112}, 10368 (2000).

\bibitem{gc} J.P. Garrahan and D. Chandler,
{\it Phys. Rev. Lett.} {\bf 89}, 035704 (2002).

\bibitem{TWBBB} C. Toninelli, M. Wyart, G. Biroli, L. Berthier,
and J.P. Bouchaud,
{\it Phys. Rev. E} {\bf 71}, 041505 (2005).

\bibitem{mayer} P. Mayer  {\it et al.},
{\it Phys. Rev. Lett.} {\bf 93}, 115701 (2004).

\bibitem{alba} G. Tarjus, D. Kivelson, S. Mossa, and C. Alba-Simionesco,
{\it J. Chem. Phys.} {\bf 120}, 6135 (2004).

\bibitem{pusey} P.N. Pusey and W. van Megen,
{\it Nature} {\bf 320}, 340 (1986).

\bibitem{hansen} J.P. Hansen and I.R. Mc Donald, 
{\it Theory of simple liquids} (Elsevier, Amsterdam, 1986).

\bibitem{lebo} J.L. Lebowitz, J.K. Percus, and L. Verlet,
{\it Phys. Rev.} {\bf 153}, 250 (1967).

\bibitem{toto}
Alternatively, this inequality may be derived from
standard probability theory. From the Cauchy-Schwarz inequality
one obtains that the average $k_{B} T^2\chi_T(t) /N=\langle \delta
C (t)\delta H (0)\rangle $ is $\le [\langle \delta
C (t)\delta C (t) \rangle \langle \delta H (0)\delta H (0)\rangle
]^{1/2}$. Using the fluctuation-dissipation relation to express
enthalpy fluctuation in terms of the specific heat at constant
pressure, one directly obtains relation \ref{ineqT}. Contrary
to the derivation in the main text, this argument does not
provide any estimate of the difference between the two sides of
the relation \ref{ineqT}.

\bibitem{djamel} D. El Masri, M. Pierno, L. Berthier, and L. Cipelletti,
{\it J. Phys. Condens. Matter} {\bf 17} S3543 (2005).

\bibitem{KA} W. Kob and H.C.  Andersen,
{\it Phys. Rev. Lett.} {\bf 73}, 1376 (1994). 

\bibitem{KAbis}
The 
numerical model is a 80:20 binary Lennard-Jones
mixture at density $\rho=1.2$.
$A$ and $B$ particles interact via a Lennard-Jones potential
$V({\bf r}_{\alpha \beta}) = 4 \epsilon_{\alpha \beta} \left[
(\sigma_{\alpha \beta}/r_{\alpha \beta})^{12} -
(\sigma_{\alpha \beta}/r_{\alpha \beta})^{6}
\right]$, with $\alpha, \beta = A, B$. Time, energy and length
are measured in units of $\sigma_{AA}$ and $\epsilon_{AA}$,
and $\sqrt{m_A \sigma_{AA}^2 / \epsilon_{AA}}$, respectively.
Other parameters are $\epsilon_{AB} = 1.5$, $\epsilon_{BB} = 0.5$,
$\sigma_{BB} = 0.88$, $\sigma_{AB} = 0.8$.
Newton equations are integrated using a velocity Verlet algorithm with
time step 0.01. Characteristic temperatures for this system are
the onset of slow dynamics, $T_o \approx 1.0$, and $T_c \approx 0.435$,
the location of the mode-coupling singularity in the analysis
of Kob and Andersen~\cite{KA}.

\bibitem{BB} G. Biroli and J.P. Bouchaud,
{\it Europhys. Lett.} {\bf 67}, 21 (2004).

\bibitem{mark} S.A. Reinsberg, X.H. Qiu, M. Wilhelm, H.W. Spiess, and
M.D. Ediger, {\it J. Chem. Phys.} {\bf 114}, 7299 (2001).

\bibitem{encoremark} X.H. Qiu and M.D. Ediger,
{\it J. Phys. Chem. B} {\bf 107}, 459 (2003).

\bibitem{WangAngell} L.M. Wang, V. Velikov, and C.A. Angell,
{\it J. Chem. Phys.} {\bf 118} 10184 (2003).

\bibitem{StevensonWolynes} J.D. Stevenson and P.G. Wolynes,
{\it J. Phys. Chem. B} {\bf 109}, 15093 (2005).

\bibitem{encoredonth} A similar correlation is discussed in \cite{tata}.


\bibitem{rfot}
X.Y. Xia and P.G. Wolynes,
{\it Proc. Natl. Acad. Sci.} {\bf 97}, 2990 (2000).

\bibitem{gc2} J.P. Garrahan and D. Chandler, 
{\it Proc. Natl. Acad. Sci.} {\bf 100}, 9710 (2003).

\bibitem{otp} S.A. Reinsberg, A. Heuer, B. Doliwa, 
H. Zimmermann, and H.W. Spiess, 
{\it J. Non-Cryst. Solids} {\bf 307-310}, 208 (2002).

\bibitem{tutu}
The energy conservation is central to this argument which does not
apply, for instance, to Langevin dynamics where the thermal bath
can locally provide energy to the system. In this case, no growing
dynamic length is expected in Arrhenius liquids.

\bibitem{tata}
E. Hempel, G. Hempel, A. Hensel, C. Schick, and E. Donth, 
{\it J. Phys. Chem. B} {\bf 104}, 2460 (2000). 

\bibitem{toto2} 
We thank A. Schofield for providing us with
the colloidal particles. L.C. is a junior member of the Institut
Universitaire de France. M.P. is supported by the MCRTN Arrested
Matter. This work was also supported by Centre National d'Etudes
Spatiales, and the French minist{\`e}re de la Recherche through an
ACI Jeunes Chercheurs.


\end{thebibliography}
\end{document}